# A zero-sum monetary system, interest rates, and implications.

Brian P. Hanley

**Abstract**

To the knowledge of the author, this is the first time it has been shown that interest rates that are extremely high by modern standards (100% and higher) are necessary within a zero-sum monetary system, and not just driven by greed. Extreme interest rates that appeared in various places and times reinforce the idea that hard money may have contributed to high rates of interest. Here a model is presented that examines the interest rate required to succeed as an investor in a zero-sum fixed quantity hard-money system. Even when the playing field is significantly tilted toward the investor, interest rates need to be much higher than expected. In a completely fair zero-sum system, an investor cannot break even without charging 100% interest. Even with a 5% advantage, an investor won't break even at 15% interest. From this it is concluded that what we consider usurious rates today are, within a hard-money system, driven by necessity.

Cryptocurrency is a novel form of hard-currency. The inability to virtualize the money creates a system close to zero-sum because of the limited supply design. Therefore, within the bounds of a cryptocurrency system that limits money creation, interest rates must rise to levels that the modern world considers usury. It is impossible, therefore, that a cryptocurrency that is not expandable could take over a modern economy and replace modern fiat currency.

**Correspondence:**  Brian P. Hanley, Butterfly Sciences, California, USA
            Email:  **brian.hanley@ieee.org**

# 1 Introduction

Examining zero-sum monetary systems is interesting in the modern world because of new cryptocurrencies that create electronic analogs of physical coin. Most of these currencies are designed to create a limited amount of electronic coins, with the idea that limited availability supports currency exchange price vis-à-vis existing currencies.

A system where interest is paid and the total amount of money is fixed is a zero-sum game – for every winner that earns interest, there must be a loser who does not, because new money is not created that allows interest or investment return payouts from any other source. The game examined here is artificial, but it models an important sector of monetary behavior – hard currency systems. Hard currency systems predate the invention of banking and they have cropped up recently in most of the crypto-currencies, of which Bitcoin is the most well-known example. This kind of monetary system has special issues that are not commonly appreciated.

Interest rates have long been a topic of concern, as high interest rates have historically been a way that the world's poor are kept in poverty. This has led to the laws against usury[1], and rules against lending for interest in some religions.

During periods when physical coin was the monetary system, the money supply was sometimes rather inflexible. With money creation either barely keeping pace with economic needs, or lagging behind, systems flirting with zero-sum money appeared, though never in pure form.

## 1.1 History

Credit preceded physical coinage in the ancient world. Credit recorded in cuneiform of ancient Sumer shows a thriving economy in which the money supply is not artificially constrained, and while accounts were kept in units of silver, silver did not actually circulate. It was, like modern money, a placeholder unit, an abstraction. Hard money coin came into being to pay standing armies, so that the ruler would not need to be the sole logistics and supply source, because coin created an instrument that could be used to pay for anything, and turned the populace into a support system. Coins were also used by governments to collect taxes sometimes, and forced the population to compete to get them so that taxes could be paid, thus enforcing the value of the currency. [2]

"Hard money" (metal standard money) appeared in roughly 600 BC, and lasted until 600 AD, when coinage declined along with a drop in slavery. The idea then reappeared and strengthened in the last 500 years with a boost from Locke in the 17$^{th}$ century[2]. Locke advised Isaac Newton to stick with gold and silver coins, even though the bullion value of the coins was higher than face value. This was rather a disaster, because some saw opportunity, for instance Scots ordered large amounts of coin from the mint, melted it down and sold it back to the mint so the mint could fulfill their next order[3]. It is curious that the mathematical genius of the age was unable to see the problem with such simple arithmetic. But Sir Isaac Newton was convinced by Locke's argument that metal money should help establish a money based on natural law.

Thus, it appears that this idea of hard money, in which coins and gold are the only "real currency" was imposed on the common man by rulers in 2 major periods. Even so, credit was created in the form of things like split tally sticks, which were exchanged as money up until 1826[4]. In the modern world, split tally sticks are similar to the practice of endorsing a check over to someone else, who in turn endorses the check to another, and so on.

So a true hard currency economy with a fixed quantity of coin is mostly fiction, even in times when hard currency coin was limited, although limitations on hard currency did create hardship, and there are reports of extreme interest rates in various countries and time periods.

## 2   Simple zero-sum investor game.

This simple version of the zero-sum investor game introduces the problem. Let us play a zero-sum game using small numbers of coins to examine the problem of interest rates. In this game there are three players. An investor, A, a borrower, B, and a competitor, C. In the game, players B and C will engage in transactions with each other. Player A provides a loan to player B, at some interest rate. After some arbitrary number of transactions, we stop the game and pay off player A. We will assume that all transaction outcomes are equally likely in this version of the game.

## 2.1   Three player investment game.

The first version of the zero-sum game will start as shown in figure 1, with 10 coins in the game. Player A has 5 coins. The other 5 coins are held by player C. Player B receives a loan from player A. We will assume initially that player B and player C are equally good at the game and that all outcomes are equally probable. This game assumes no carryover from one round to the next. Each round of some arbitrary number of transactions starts fresh.

Player A loans his 5 coins to player B with simple interest of 1 coin (20%) to be paid at the end of each round of the game. Player A waits while player B transacts with player C.

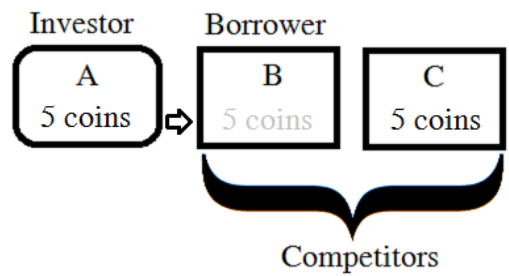

*Figure 1*: Simple 3 player game

If Player B makes at least a coin of net income, it must come from player C. Conversely, if player C nets a coin, that will be at the expense of player B. Over half of the ways this game can turn out have player B losing or breaking even. This is true because there is only one break-even state, and the rest are evenly divided between wins and losses for player B.

Player A has a serious problem. Player A has 5 ways to lose money, 1 way to make nothing, and only 5 ways make money – but he only makes one coin each time. So, if player A has 10 iterations of the game, and outcomes are evenly distributed, then he must lose overall, as shown in Table 1.

*Table 1*: Zero-sum coin game with 1 coin interest for 5 coins loaned.

| Game Iteration | Capital loaned to Player B | Game outcomes Player B | Game outcomes Player C | Player B Net | Player A Win | Player A Loss | Player A Net |
|---|---|---|---|---|---|---|---|
| 1 | 5 | 0 | 10 | 0 |  | 5 | 0 |
| 2 | 5 | 1 | 9 | 0 |  | 4 | 1 |
| 3 | 5 | 2 | 8 | 0 |  | 3 | 2 |
| 4 | 5 | 3 | 7 | 0 |  | 2 | 3 |
| 5 | 5 | 4 | 6 | 0 |  | 1 | 4 |
| 6 | 5 | 5 | 5 | 0 | 0 | 0 | 5 |
| 7 | 5 | 6 | 4 | 0 | 1 | 0 | 6 |
| 8 | 5 | 7 | 3 | 1 | 1 | 0 | 6 |
| 9 | 5 | 8 | 2 | 2 | 1 | 0 | 6 |
| 10 | 5 | 9 | 1 | 3 | 1 | 0 | 6 |
| 11 | 5 | 10 | 0 | 4 | 1 | 0 | 6 |
| Totals |  |  |  | 10 | **5** | **15** |  |

So, assuming both players are equally good, and outcomes distribute evenly, for every 5 coins player A can get in interest, he should lose 15 coins. What rational player would loan money at 20% (e.g. 1 coin on a 5 coin loan) interest under these circumstances? He is guaranteed to lose 3 times for every win. However, player B, the borrower, has net positive outcome. It is a good deal for him.

What interest rate is needed for player A to break even on his investments? The number is much higher than most would guess. Player A has to charge 100% interest to break even in this game as seen in Table 2.

*Table 2*: Coin game with 100% interest rate, 5 coins interest for 5 coins loaned.

| Game Iteration | Player A Capital loaned to Player B | Game outcomes Player B | Game outcomes Player C | Player B Net | Player A Win | Player A Loss | Player A Net |
|---|---|---|---|---|---|---|---|
| 1 | 5 | 10 | 0 | 0 | 0 | 5 | 0 |
| 2 | 5 | 9 | 1 | 0 | 0 | 4 | 1 |
| 3 | 5 | 8 | 2 | 0 | 0 | 3 | 2 |
| 4 | 5 | 7 | 3 | 0 | 0 | 2 | 3 |
| 5 | 5 | 6 | 4 | 0 | 0 | 1 | 4 |
| 6 | 5 | 5 | 5 | 0 | 0 | 0 | 5 |
| 7 | 5 | 4 | 6 | 0 | 1 | 0 | 6 |
| 8 | 5 | 3 | 7 | 0 | 2 | 0 | 7 |
| 9 | 5 | 2 | 8 | 0 | 3 | 0 | 8 |
| 10 | 5 | 1 | 9 | 0 | 4 | 0 | 9 |
| 11 | 5 | 0 | 10 | 0 | 5 | 0 | 10 |
| Totals | | | | 0 | **15** | **15** | |

The simple interest has to be over 100% for Player A to average simple breakeven, if both players start even. For player A, the investor, to win more than he loses over a large number of rounds, the game must be skewed by having more upside possibility than downside.

### 2.2   Multi-player game using Gaussian distribution.

Now let us look at a more realistic case with many players in the game. To do this we will add the assumption that outcomes of many transactions of many players will follow a Gaussian distribution that averages to zero. This model assumes that all players transact freely, without hoarding, and that players include three classes that correspond to players A, B and C.

*Figure 2*: Gaussian weighted investment risk graph

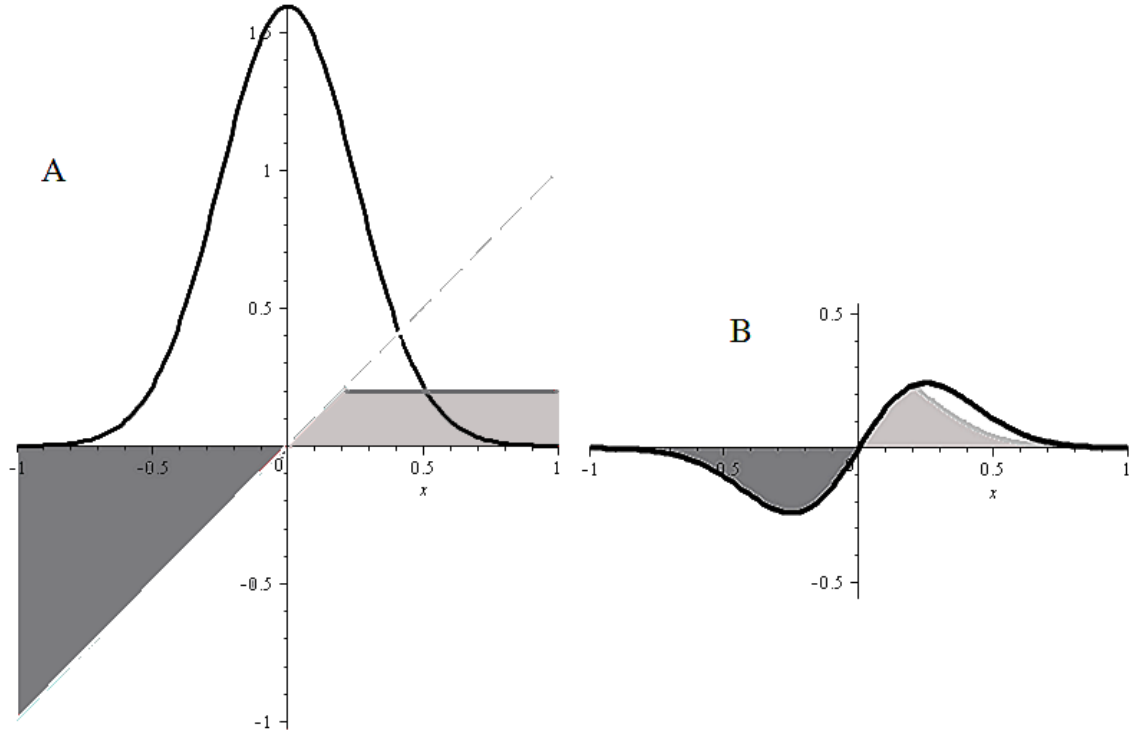

**2A:** Gaussian distribution overlaid on linear plot of table 1. *Black line* is Gaussian distribution with *σ* = 0.25, and *μ* = 0. *Gray dashed line* shows zero-sum risk/reward per table 2. *Solid gray line* shows win limited to 20% over capital risked. Zero average (*μ*) of the Gaussian distribution represents the odds of win or loss in any round between players are, on average, even.

**2B:** Combined equation 1 shows plot of Gaussian distribution and win/loss outcomes. Black line is zero-sum risk of win vs. loss with no limit on investor return, up to the full amount of the investment. Gray line shows win limited to 20% gain.

***Dark gray region*** in both graphs is area of wins for player C. Light gray region is the area of wins for player A.

$$\frac{1}{\sigma\sqrt{2\pi}} e^{-\frac{(x-\mu)^2}{2\sigma^2}} \tag{1}$$

**Equation 1:** Gaussian black line in figure 2B. *Where: x = loss or gain in the game.*

$$I\left(\frac{1}{\sigma\sqrt{2\pi}}e^{-\frac{(x-\mu)^2}{2\sigma^2}}\right) \quad (2)$$

**Equation 2:** Gray line in figure 2B. *Where: $I$ = simple interest, $\sigma$ = standard deviation, $\mu$ = mean/median/mode. Values: $\sigma$ = 0.25, $I$ = 0.2, and $\mu$ = 0.*

The value of 0.25 was selected for $\sigma$ because this produces an equation that allows us to restrict our range between -1 and 1 and the elements beyond the range have virtually no contribution.

### 2.3　Multi-player Gaussian game where $\mu$ varies in favor of player class B.

*Figure 3*: Gaussian weighted investment risk graph

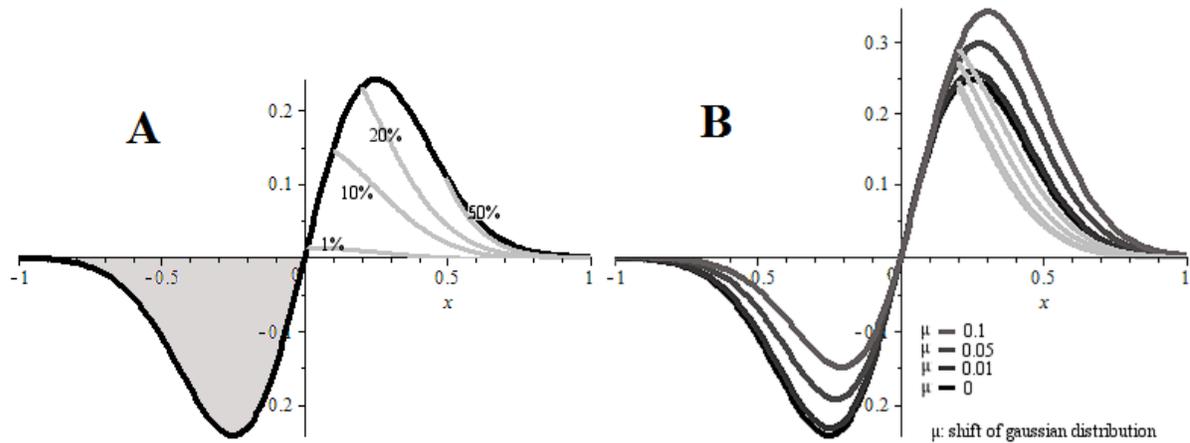

**3A:** Equations 1 & 2 where $I$ rate varies: 1%, 10%, 20%, and 50%. $\sigma$ = 0.25, $\mu$ = 0.
**3B:** Equations 1 & 2 where $\mu$ varies: 0, 0.01, 0.05, 0.1. $\sigma$ = 0.25, $I$ = 20%. If the area to the right of the y axis up to the light gray curve is larger than the area shown in gray to the left of the y axis then player A can consistently win.

Inspecting figure 3, one can see that even at 20% interest, by shifting the center of the Gaussian curve to the right, the area under the interest curve becomes larger than the area of wins for player C.

From equation 2, an equation that can tell us what the average fraction of capital loss will be for a particular simple interest can be developed. Equation 3 represents dividing the area of the gray region of figure 2B above zero (win) by the area of figure 2B below zero (loss). Thus, we integrate over the region of the curves, and adjust so that anything below one is a negative return, a loss. Equation 3 replaces individual players by classes of players in each category.

$$\frac{\left(\int_0^I x\left(\frac{1}{\sigma\sqrt{2\pi}}e^{-\frac{(x-\mu)^2}{2\sigma^2}}\right)dx + \int_I^1 I\left(\frac{1}{\sigma\sqrt{2\pi}}e^{-\frac{(x-\mu)^2}{2\sigma^2}}\right)dx\right)}{\left|\int_{-1}^0 x\left(\frac{1}{\sigma\sqrt{2\pi}}e^{-\frac{(x-\mu)^2}{2\sigma^2}}\right)dx\right|} - 1 \quad (3)$$

**Equation 3:** *Where: **I** = Simple interest, **σ** = standard deviation, **μ** = mean/median/mode.*

Using Equation 3, we can see what average win/loss is expected for an interest rate and the transaction unfairness. If the scales are unfairly tilted in favour of the player class B that is backed by our investor class A, then investor class A can net a positive return. By shifting the distribution to the positive side by rigging the game's odds, an investor can make money. Even so, this can only occur up to the limit of the money in the game. The ability of an investor to win at this game requires a higher level of interest than one might expect, even with significant "thumb on the scales" tilting to advantage the investor.

*Figure 4*: Gaussian loss versus return on zero-sum investment

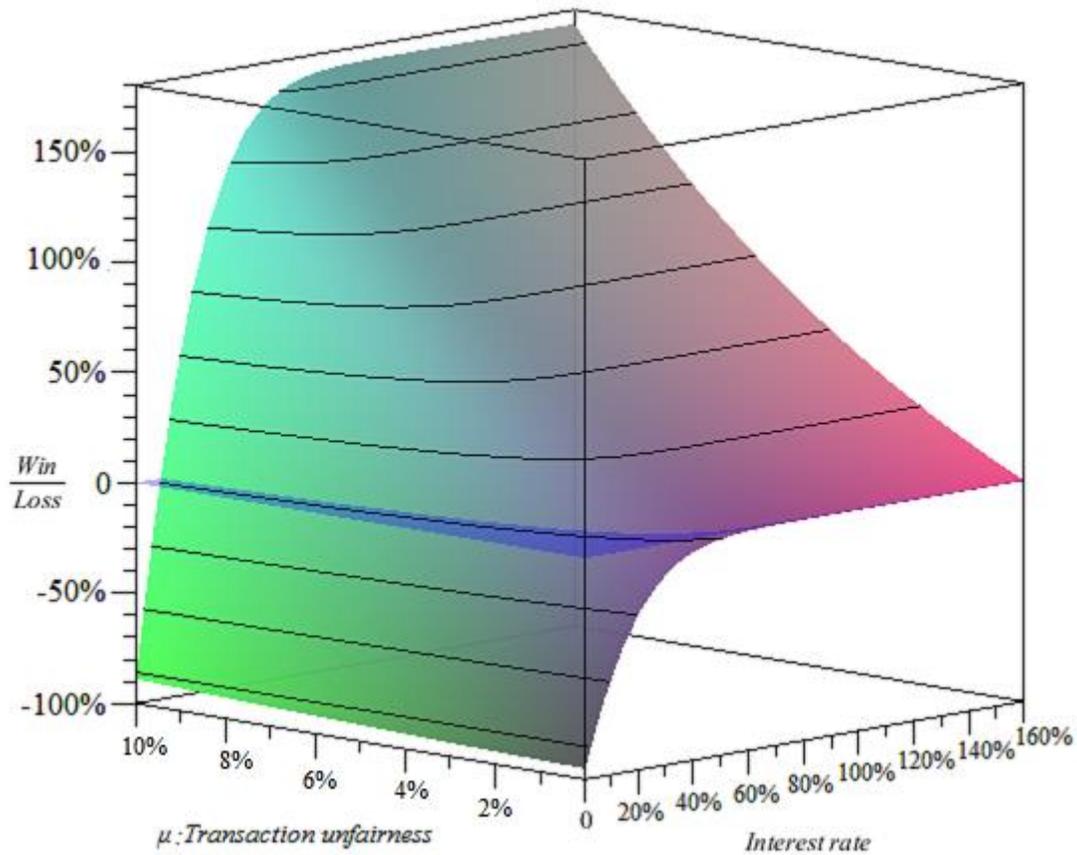

$\mu$ = 0 to 10%, $\sigma$ = 0.25, $I$ = 1% to 160%.

Computations performed using Maple™[5].

## 3    Discussion

Without a mechanism to make credit available as needed, or the ability to otherwise expand the money supply in concert with economic activity as China did by purchasing the gold and silver from the new world[2], interest payments can only cannibalize other lending to pay winners. There must be a loss for every gain earned in interest.

This conundrum means that in a zero-sum game, a rational player rigs the game, or else charges outrageous amounts of interest/investment return. History has recorded that both responses to investment risk have been common. Historical accounts of interest rates far exceeding 100% exist, and often it is presumed that these interest rates were necessarily evidence of greed without rationality.

Even in a zero-sum game in which rationality and understanding is not assumed, repeated iterations should evolve players who play the game according to winner's rules. We can see suggestions of this in past centuries in which usury was regularly attacked and social rules made to control it. For instance, the ancient Jewish temple system had periodic jubilees when all debts were forgiven[6]. The temple received tithes, so its money supply was consistently replenished, allowing it to make hard currency loans. Tithes allowed loan activity to continue in a zero-sum system that did not charge excessive interest. The Babylonians had a similar tradition of hubullum masa'um (washing away debt), Assyrians left records of debt forgiveness, as did many other rulers and dynasties of ancient Mesopotamia[7].

Ancient Rome and Greece had common nominal rates on the order of 1% per month, and sometimes loans were available as low as 4% per year. The Roman Empire, in its peak, had nominal rates on the order of 1% per month, although rates exceeding the legal limit of 12% occurred frequently[8].

How ancient Rome could have operated a presumed hard-money system with such formally low interest rates can be looked at in several ways. First, there may have been a significant difference between the nominal official interest rate in Rome and what ordinary borrowers could get access to, just as there is today in our banking system, but with larger differentials. There is evidence of this in Rome, Greece and England. In Julius Caesar's time, Brutus charged 48%, rates as high as 48% per month (10,900% per year) are recorded in ancient Athens, and interest of 52%-120% per year was charged in 12$^{th}$ century England[9]. Such rates are more consistent with the zero-sum game discussed here.

Second, the Roman Empire was expanding for centuries, taking in new treasure from conquered regions. For many years Greece also plundered, bringing back booty and slaves, much as the Vikings did in Europe centuries later. With plunder providing increases to the money supply, low cost slave labor, mining of precious metals, and mostly restricting money to the citizen elite and an upper class of slaves, the money supply could meet or exceed needs for long periods. This could create periods in which an excess of hard-money obtained by plunder and seignorage would allow periods of low interest rates in the core of the empire because, in effect, the money supply could be managed for the elite.

Third, the wealthiest, most powerful class could afford to lose significant amounts of money because the society was rigged to supply them with more. Private parties with high levels of wealth could thus have been tied into the plunder system of various empires.

Finally, there is evidence that deposits existed in Roman times, and perhaps some degree of banking-like transactions[9]. It is plausible that the money supply due to money creation in the modern sense may have been somewhat larger from time to time than the hard-money supply because of the use of accountants and letters to exchange large amounts of money that were inconvenient to transport physically.

## 4    Conclusion

The interest requirements for investors in limited hard-money systems are considerably higher than most realize. Without corruption to rig the outcome, it is impossible for an investor class to make money in a hard-money system that does not have regular, managed, inflow of hard currency. Within that investor class, stochastic variation will create winners and losers in a fair system, but the class as a whole will not. Only within hard money systems that harvest a regular supply of new commodity money (e.g. gold, silver, gemstones or other commodity money) and manage the release of that money into their hard-currency system can interest rates fall below the 100% level and reward investors as a class.

Cryptocurrencies are neo-coin systems created within the context of a more sophisticated monetary mechanism of fiat currency and money creation by loans. The inability to virtualize money prevents creation of new money by loans, so the cryptocurrency system itself becomes zero-sum.  Therefore, within the bounds of a cryptocurrency system that limits money creation, interest rates must rise to levels that the modern world considers usury. Consequently, it is impossible for such a system of electronic coinage to expand and become the primary currency of any modern economy.